\documentclass[aip,pof]{revtex4-1}

\usepackage{graphicx}
\usepackage{dcolumn}
\usepackage{bm}
\usepackage{psfrag}
\usepackage[english]{babel}

\begin{document}

\preprint{APS/123-QED}

\title{Finger properties in bounded double diffusive finger convection}

\author{A. Rosenthal}

\author{A. Tilgner}
\email[corresponding author: ]{andreas.tilgner@phys.uni-goettingen.de}

\affiliation{Institute of Astrophysics and Geophysics, University of G\"ottingen,
Friedrich-Hund-Platz 1, 37077 G\"ottingen, Germany }

\date{\today}

\begin{abstract}
We analyze experimental data on double diffusive convection in an
electrochemical cell in the finger regime. All fingers in the experiments are
bounded on at least one end by a solid wall. The properties of these fingers are
compared with those of fingers in other experiments which are surrounded by
fluid on all sides. The compositional boundary layers are found to be thinner
than the finger width. The finger thickness agrees well with half the wavelength
of the fastest growing mode obtained in linear stability analysis. The ion transport
through the boundary layers is reduced by two orders of magnitude compared with
unbounded fingers. The overturning layers in staircases contribute negligibly to
salinity mixing because of efficient transport between finger layers and convection
rolls.
\end{abstract}

\maketitle

\section{Introduction}
Double diffusive convection occurs if the density of a fluid is governed by two
different properties with different diffusion rates, such as temperature and
salinity in ocean water. Ocean water is the paradigmatic example of a fluid
prone to double diffusive effects, but many other examples exist for this
phenomenon \cite{Turner74, Turner85, Radko05}. The behaviour of double diffusive
convection very much depends on which property acts to destabilize the fluid. In
this paper, we will be concerned with the case in which the rapidly diffusing
component (temperature in the example of sea water) on its own leads to a stable
stratification of the medium while the slowly diffusing property (salinity in the
case of sea water) drives an overturning motion. This combination of temperature
and salinity gradients may lead to finger convection in which the convection
cells take the form of slender cells elongated along the vertical. Depending on
control parameters, these fingers occupy the entire fluid volume or they occur
in vertically stacked layers separated by layers filled with wide convection
rolls familiar from simple Rayleigh-B\'enard convection. Flows of the second
kind are called staircases.

This type of convection already was the topic of numerous numerical and
experimental studies \cite{Radko05}. The experiments \cite{Tilgne24} are limited
in the parameter range they can explore by the available fluids and the size of
the apparatus. They also face the challenge of maintaining a sufficiently
constant salinity gradient. Most experiments actually investigate non
stationary conditions in which an initial salinity gradient or a salinity
difference between two layers is eroded in the course of time due to convection.
In these systems, fingers are similar to fingers in an unbounded domain in that
they end within the fluid domain and are surrounded by fluid from all sides.
Those experiments that maintain truly stationary conditions 
\cite{Krishn03, Krishn09, Hage10} generate fingers at
least some of which are limited in their vertical extent by solid boundaries.
Most numerical simulations also investigated unbounded fingers and only few
numerical studies of bounded fingers at high Rayleigh numbers
exist \cite{Yang15, Yang16_b, Yang20, Li23}.

The present paper analyzes data obtained from
experiments on double diffusive electrochemical convection. As opposed to other
applications of electrochemical convection \cite{Mani20}, care was taken to
eliminate the effects of electrical fields so that the dynamics are solely
determined by advection and diffusion. The important characteristics of this
experiment in comparison with other double diffusive systems is that the system
can be maintained in a steady state and the control parameters describing
material properties are large. The Prandtl and Schmidt numbers (to be defined
below, and see table \ref{table:nomenclature} for a summary of the nomenclature) 
are around 9 and 2000, respectively, whereas they are 7 and 700 for sea
water. The fingers observed in the experiments analyzed here have at least one
tip bordering a solid plate, so that the effect of a solid boundary can be
investigated through comparison with results on unbounded fingers.

The next section reviews the experimental procedures. Previous publications on
double diffusive electrochemical convection \cite{Hage10, Kellne14, Rosent22}
focused on transitions between different flow regimes which could not be mapped
out in other experiments. The aim of the present paper is to connect
observations on finger properties in double diffusive electrochemical convection
with analogous results on different double diffusive systems. These concern the
finger thickness in the next section, and the salinity transport through finger
layers and staircases in sections \ref{section:flux_fingers} and
\ref{section:flux_staircases}, respectively.

\begin{table}[b]
\centering
\begin{tabular}{|c|c|c|}
\hline

$\nu$, $\kappa$, $D$ & diffusion constants for momentum, temperature and salinity & $m^2 s^{-1}$ \\
\hline

$g$ & gravitational acceleration & $m s^{-2}$ \\
\hline

$\alpha$ & thermal expansion coefficient & $K^{-1}$ \\
\hline

$\beta$ & compositional expansion coefficient & $mol^{-1}$ \\
\hline

$L$ & height of the cell & $m$ \\
\hline

$L_f$ & height of a finger layer& $m$ \\
\hline

$d$ & finger thickness & $m$ \\
\hline

$\lambda_c$ & thickness of compositional boundary layer & $m$ \\
\hline

$T$ & temperature & $K$ \\
\hline

$\theta$ & difference between the temperature and the linear conduction profile & $K$ \\
\hline

$\Delta T$ & temperature difference between bottom and top of the cell& $K$ \\
\hline

$\Delta T_f$ & temperature difference between bottom and top of a finger layer& $K$ \\
\hline

$c$ & concentration & $mol/m^3$ \\
\hline

$s$ & difference between the concentration and the linear diffusion profile & $mol/m^3$ \\
\hline

$\Delta c$ & concentration difference between top and bottom of the cell& $mol/m^3$ \\
\hline

$\Delta c_f$ & concentration difference between top and bottom of a finger layer& $mol/m^3$ \\
\hline

$c_0$ & concentration averaged over the cell & $mol/m^3$ \\
\hline

$F$ & Faraday constant & $A \, s \, mol^{-1}$ \\
\hline

$F_s$ & salinity flux& $mol \, m^{-2} s^{-1}$ \\
\hline

$j$ & current density & $A m^{-2}$ \\
\hline

$Sh$ & Sherwood number &  \\
\hline

$\Lambda$ & density ratio &  \\
\hline

$Ra_T$, $Ra_c$, $Re$ & thermal Rayleigh, compositional Rayleigh and Reynolds numbers &  \\
 & based on the height of the cell &  \\
\hline

$Ra_{T,f}$, $Ra_{c,f}$, $Re_f$ & thermal Rayleigh, compositional Rayleigh and Reynolds numbers&  \\
 & based on the height of a finger layer&  \\
\hline

$Pr$ & Prandtl number &  \\
\hline

$Sc$ & Schmidt number &  \\
\hline

\end{tabular}

\caption{Most symbols occur only locally in this paper and are defined where they are
used. Symbols that occur throughout the paper are collected in this table. The third column
is empty for non dimensional variables and contains the units otherwise.}
\label{table:nomenclature}
\end{table}

\begin{figure}
\includegraphics[height=10cm]{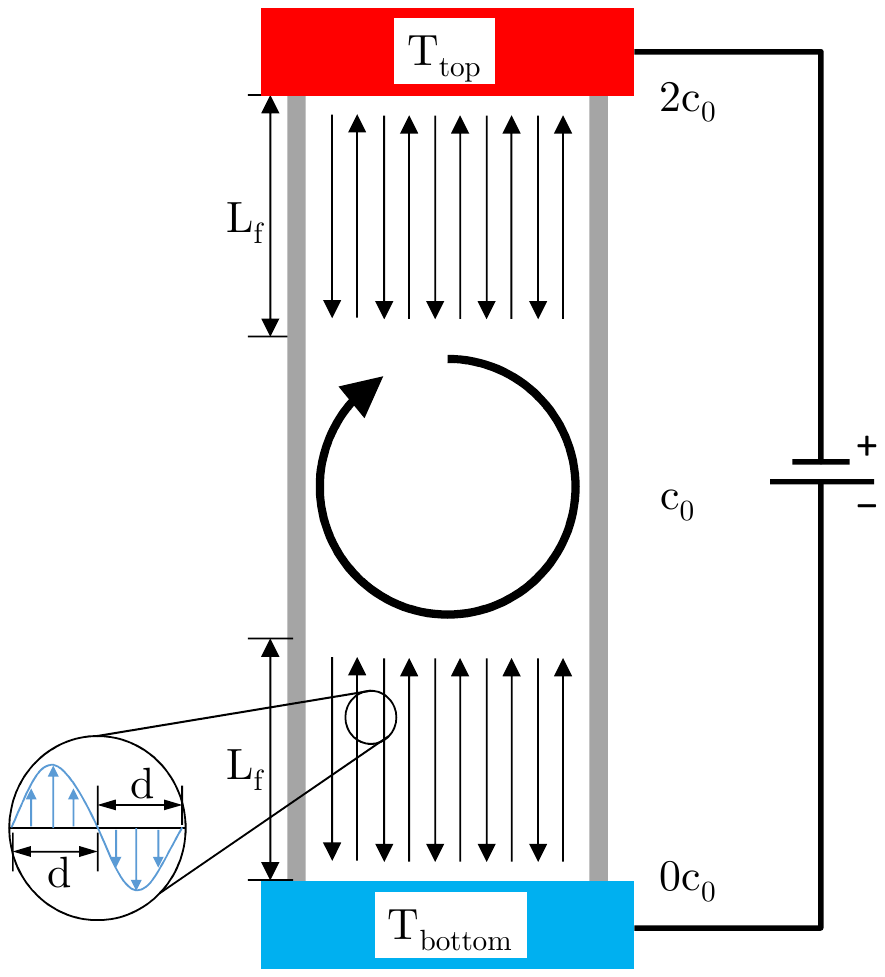}
\caption{Sketch of the experiment. Two copper electrodes are regulated in temperature, the upper
electrode is heated while the lower electrode is cooled. The average concentration of copper ions
in the electrolyte is $c_0$. An applied voltage drives the limiting current through the electrolyte
so that the copper concentration is zero at the bottom electrode and $2 c_0$ at the top electrode.
Fingers of thickness $d$ either cross the entire cell or (as sketched in this figure)
they are part of a staircase in which case
they fill two layers of height $L_f$ each with a layer with overturning motion in between the two
finger layers. The velocity field within fingers is symbolically sketched in the magnified excerpt.}
\label{fig:sketch}
\end{figure}

\section{The Experiment}
\label{Experiment}

The electrochemical system under study sketched in Fig. \ref{fig:sketch}
is introduced in detail in Ref.  \onlinecite{Goldst90} and its use for 
double diffusive convection is explained in
Refs. \onlinecite{Hage10, Kellne14, Rosent22} so that only a brief summary of the technique
is presented here. The convecting fluid is a solution of cupric sulfate in
sulfuric acid contained in a cell with top and bottom boundaries made of copper.
These two boundaries are regulated in temperature to maintain a temperature
difference from top to bottom. In addition, a voltage is applied to the two
boundaries which then act as electrodes with copper ions detaching at the top
and reattaching at the bottom boundary. The ions of the acid do not participate
in the electrochemistry and accumulate in microscopic layers next to the
boundaries so that the bulk of the cell is free of electric fields and the
copper ions diffuse and are advected the same as salinity in other double
diffusive systems.

Four control parameters describe the system within the Boussinesq approximation.
The Prandtl and Schmidt numbers,  $Pr$ and $Sc$, are defined as
\begin{equation}
Pr=\frac{\nu}{\kappa} ~~~,~~~ Sc=\frac{\nu}{D},
\end{equation}
with $\nu$ the kinematic viscosity, $\kappa$ the thermal diffusivity, and $D$
the diffusion constant of copper ions. The other two control parameters are the
thermal and chemical Rayleigh numbers,  $Ra_T$ and $Ra_c$, given by
\begin{equation}
Ra_T=\frac{g \alpha \Delta T L^3}{\kappa \nu}  ~~~,~~~ 
Ra_c=\frac{g \beta \Delta c L^3}{D \nu},
\end{equation}
where $g$ is the gravitational acceleration, $L$ is the height of the cell,
$\alpha$ and $\beta$ are the thermal and compositional expansion coefficients,
and $\Delta T$ and $\Delta c$ are the temperature and copper ion concentration
differences between top and bottom. The sign convention is chosen such that
$\alpha$ and $\beta$ are positive and 
\begin{equation}
\Delta T = T_{\rm{bottom}}-T_{\rm{top}} ~~~,~~~ \Delta c = c_{\rm{top}}-c_{\rm
{bottom}}
\end{equation}
so that positive Rayleigh numbers correspond to unstable and negative Rayleigh
numbers to stable stratification.

The Prandtl and Schmidt numbers are essentially fixed in our experiments at 
$Pr \approx 9$ and $Sc \approx 2000$. The Rayleigh numbers on the other hand can
be varied, most efficiently by using cells of different heights. Values for $L$
of $2 cm$, $4 cm$, $8 cm$, $20 cm$, $40 cm$ and $80 cm$ appear in our data set.
The $\Delta T$ in $Ra_T$ is conveniently adjusted through the temperature
control of the electrodes, while $\Delta c$ in $Ra_c$ is varied by using
solutions of different copper concentration (or ''salinity'') $c_0$. The cell is
always operated at the limiting current \cite{Hage10,Goldst90} which implies
that $\Delta c = 2 c_0$. 

It will sometimes be useful to replace one of the Rayleigh numbers by another
control parameter which is the density ratio $\Lambda$ given by
\begin{equation}
\Lambda=\frac{Ra_T}{Ra_c} \frac{Sc}{Pr} = \frac{\alpha \Delta T}{\beta \Delta
c}.
\end{equation}

\begin{figure}
\includegraphics[width=10cm]{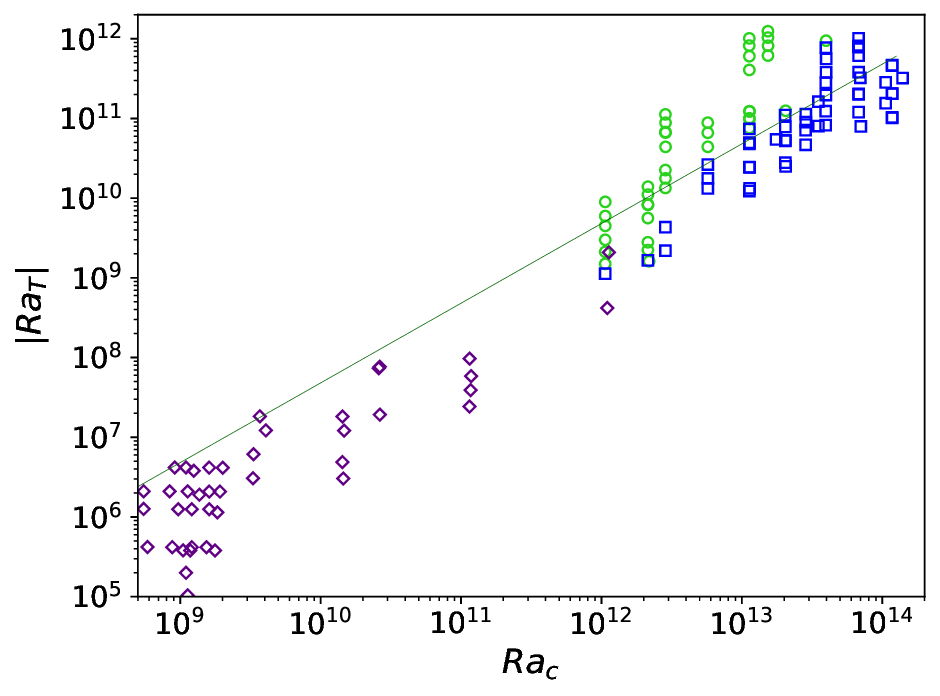}
\caption{Overview of the available experimental data in the ($Ra_c,Ra_T$)-plane. 
Circles and diamonds represent states with fingers
crossing the entire cell from top to bottom whereas squares are staircases. 
The straight line indicates $|\Lambda|=1$.}
\label{fig:param_space}
\end{figure}

The data analyzed in this paper is taken from the
experiments described in Refs.  \onlinecite{Hage10,Rosent22}. The Rayleigh numbers of
the available data are shown in Fig. \ref{fig:param_space}.
The data points are split into three classes in this figure: 
They are either taken from Ref. \onlinecite{Hage10} (indicated by diamonds) and
correspond to flow states with fingers extending from one boundary to the other
at $Ra_c$ small enough (except for two points) so that staircases do not appear 
at any $|Ra_T|$ for this $Ra_c$, or to flow states also with
fingers crossing the entire cell but at $Ra_c$ large enough so that staircases
form at suitable $|Ra_T|$ (circles in Fig. \ref{fig:param_space}), and finally
the staircases themselves in which the fingers are interrupted by a convection
roll (shown by squares).  These symbols will be used again in the subsequent
figures because it facilitates their reading and conveniently identifies the
origin of the data points. A few crude features are associated with the
different types of points. The diamonds are flow states with very regular fingers
and nearly all these points have $|\Lambda|<1$. The data collected at these
Rayleigh numbers obey simple scaling laws \cite{Hage10}. The circles correspond
to more diverse flow states \cite{Rosent22}. The fingers at high $Ra_c$ are more
time dependent and disordered, sometimes with finger widths varying as a
function of height. The staircases all
have the same structure: They consist of a top and bottom finger layer with an
intervening overturning layer. No staircases with fewer or more layers appear in
the data set.

The most obvious measurement collected from the experiments is the Sherwood
number $Sh$ which quantifies the ion flux through the layer and which is defined
as
\begin{equation}
    Sh=\frac{j\,L}{z\,F\,D\,\Delta c}
\label{eq:def_Sh}
\end{equation}
if $j$ is the current density, $z$ the valence of the ion ($z=2$ for $Cu^{2+}$),
and $F$ is Faraday's constant. Sherwood number is the usual designation for the
non dimensional form of the ion flux through the cell within the electrochemical
literature, whereas it is typically called the chemical Nusselt number or the
salinity Nusselt number in the literature dedicated to double diffusion.

The velocity field is obtained from particle image velocimetry (PIV) in a
vertical plane near the center of the cell. PIV allows us to determine one
horizontal component of velocity $v_x$ together with the vertical component 
$v_z$. An average velocity $v$ is then computed as 
\begin{equation}
v = \left( \frac{1}{A} \int \langle v_x^2+v_z^2  \rangle dA \right)^{1/2}
\end{equation}
where the integral extends over the area $A$ of the vertical plane containing
fingers. This is the area of the entire plane illuminated for PIV in the case of
fingers crossing the whole cell, but it excludes the area filled by wide
convection rolls in staircases. The angular brackets denote average over time.
The velocity $v$ is then used to compute a global Reynolds number $Re$, and when
dealing with staircases, it is also useful to define a Reynolds number on the
scale of the finger height $L_f$:
\begin{equation}
Re = \frac{v L}{\nu}   ~~~,~~~~
Re_f = \frac{v L_f}{\nu}.
\end{equation}
The thickness of the finger layers $L_f$ is extracted from the velocity fields
obtained by PIV \cite{Rosent22}. One would also like to attribute Rayleigh
numbers to finger layers within staircases, but we do not know the concentration
and temperature differences between the top and bottom of these layers.
Let us first define the Rayleigh numbers $Ra_{T,f}$ and $Ra_{c,f}$ of finger layers as
\begin{equation}
Ra_{T,f}=\frac{g \alpha \Delta T_f  L_f^3}{\kappa \nu}  ~~~,~~~ 
Ra_{c,f}=\frac{g \beta \Delta c_f L_f^3}{D \nu}.
\label{eq:Ra_f}
\end{equation}
It will be argued below that the following definitions for $\Delta T_f$, $\Delta
c_f$ and also $L_f$ are reasonable in figures showing data for both staircases
and fingers crossing the whole cell: If fingers extend from one boundary to the
other, then one obviously selects $L_f=L$, $\Delta T_f=\Delta T$ and
$\Delta c_f=\Delta c$. For finger layers in staircases (which all contain two
finger layers in our data base), $L_f$ is the layer
depth, $\Delta T_f=\Delta T/2$ and $\Delta c_f=\Delta c$. This last definition
of $\Delta c_f$ is different from the one used in Ref. \onlinecite{Rosent22}.
Section \ref{section:flux_staircases} will explain why this modification is
necessary.

There is a characteristic length scale that can be extracted from the PIV
measurements and that we will identify with a finger thickness. It is
straightforward to determine a reasonable finger thickness for the very regular
time independent flows at low $Ra_c$, but a more refined treatment is necessary
at high $Ra_c$. A definitive characterization of finger shape requires 3D
velocity fields which are not available. \citet{Hage10} concluded from the
combination of PIV and Sherwood number measurements that the fingers are
lamellar rather than tubular. Different orientations of the lamellae with
respect to the plane of illumination of the PIV lead to different apparent
finger thicknesses. The minimal thickness is observed if lamellae and plane of
illumination are orthogonal to each other, and the thickness measured in this
geometry is what one would call finger thickness if horizontal cross sections of
the velocity field were available. However, one has to find a compromise between
the search for a minimum and the need for noise reduction through averaging, so
that we opted for the following procedure: First, count in every instantaneous
velocity field obtained from PIV the number of zero crossings of $v_z$ as a
function of the horizontal coordinate $x$ and deduce an average separation of
zeros of $v_z$ at every height $z$. Then, select the height at which this
average is minimal. This is also done with flow states in mind which are near
the transition form pure finger convection to staircases. When a central layer
with a broad convection cell develops, the fingers widen near the interface to
the convection roll \cite{Rosent22}, and we assume the finger size far away from that interface
to be the relevant finger size. Finally, the minimal thickness thus obtained in every instantaneous 
velocity field is averaged over time. This yields the length $d$ which we use
for finger thickness in the data analysis below.

\begin{figure}
\includegraphics[width=10cm]{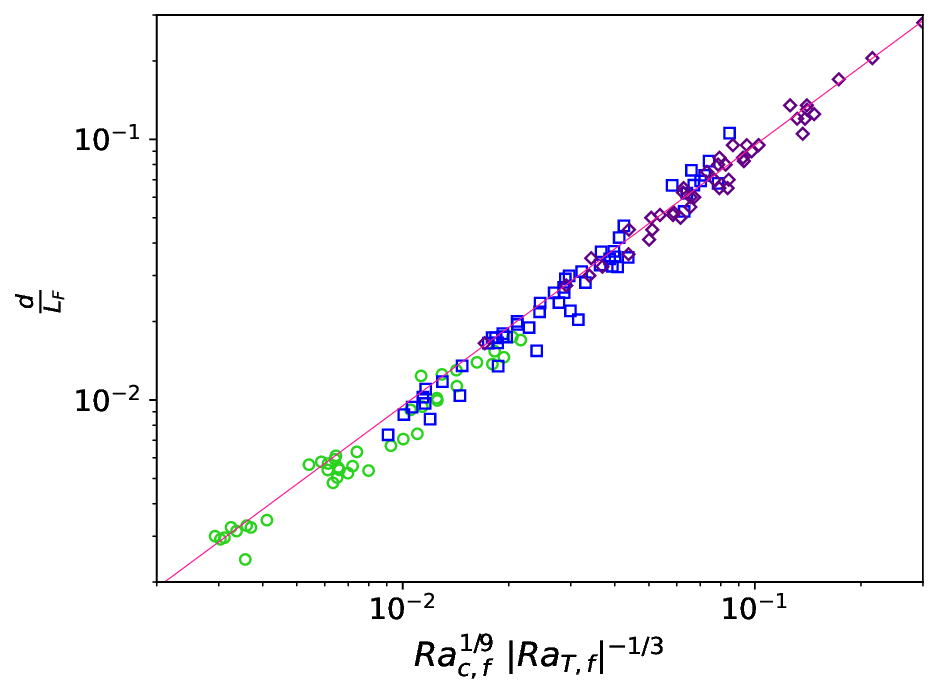}
\caption{The finger width $d$ divided by the finger height $L_f$ as a function
of $|Ra_{T,f}|^{-1/3} Ra_{c,f}^{1/9}$ to verify eq. (\ref{eq:d_Ra}) which is
represented by the continuous line. The symbols are the same as in fig.
\ref{fig:param_space}.}
\label{fig:d_L}
\end{figure}

\section{Finger thickness}
\label{section:thickness}

The finger thickness is a quantity that received much interest from both
numerical simulations and experiments because there is a simple hypothesis to
test: It is tempting to assume that for any given stratification, the finger
mode with the fastest growth rate according to linear stability analysis contains
fingers of the size that will be observed in the saturated non linear state.
This section will compare the experimentally obtained finger thickness with the
predictions deduced from this assumption.

The experiments on the regular fingers at low $Ra_c$ obey a simple scaling law
\cite{Hage10}. If we generalize the variables to include fingers in staircases,
this scaling law is 
\begin{equation}
\frac{d}{L_f} = 0.95 |Ra_{T,f}|^{-1/3} Ra_{c,f}^{1/9}.
\label{eq:d_Ra}
\end{equation}
As Fig. \ref{fig:d_L} shows, this relation fits very well all types of fingers,
which corroborates the scaling law, but also the procedure to determine finger
widths and the definitions of $Ra_{T,f}$ and $Ra_{c,f}$. However, the relation
with fastest growing modes is not apparent.

The equations of evolution for the fields of velocity $\bm v(\bm r,t)$, temperature
$T(\bm r,t)$, salinity or concentration $c(\bm r,t)$ and pressure $p(\bm r,t)$
are given by
\begin{equation}
\partial_t \bm v + (\bm v \cdot \nabla) \bm v 
= - \frac{1}{\rho} \nabla p + \nu \nabla^2 \bm v + g \alpha T \hat {\bm z} 
-g \beta c \hat {\bm z} 
\label{eq:NSE}
\end{equation}
\begin{equation}
\nabla \cdot \bm v = 0
\end{equation}
\begin{equation}
\partial_t T + \bm v \cdot \nabla T = \kappa \nabla^2 T 
\label{eq:T}
\end{equation}
\begin{equation}
\partial_t c + \bm v \cdot \nabla c = D \nabla^2 c 
\label{eq:c}
\end{equation}
where $\hat {\bm z}$ denotes a unit vector in the vertical direction, $\rho$ is
the density and the other variables have the meaning assigned to them in the
previous section. The lower and upper boundaries are represented by the
conditions that $\bm v = 0$ and that $T$ and $c$ have prescribed fixed values on the
boundaries. However, theoretical analysis is easiest in an infinitely extended
domain. We therefore split $T$ and $c$ into horizontal and temporal averages 
$\overline{T}$ and $\overline{c}$ and fluctuations $\theta$ and $s$ around that
mean:
\begin{equation}
T(\bm r,t) = \overline{T}(z) + \theta(\bm r,t)
\end{equation}
\begin{equation}
c(\bm r,t) = \overline{c}(z) + s(\bm r,t).
\end{equation}
In a plane layer with $T$ and $c$ imposed on horizontal boundaries, $\overline{T}$ and $\overline{c}$
are a priori unknown and have to be determined as part of the solution, whereas
they are imposed as the driving agent in models in infinitely extended volumes,
in which case it is convenient to choose $\overline{T}$ and $\overline{c}$ such
that their gradients are constant.

\begin{figure}
\includegraphics[width=10cm]{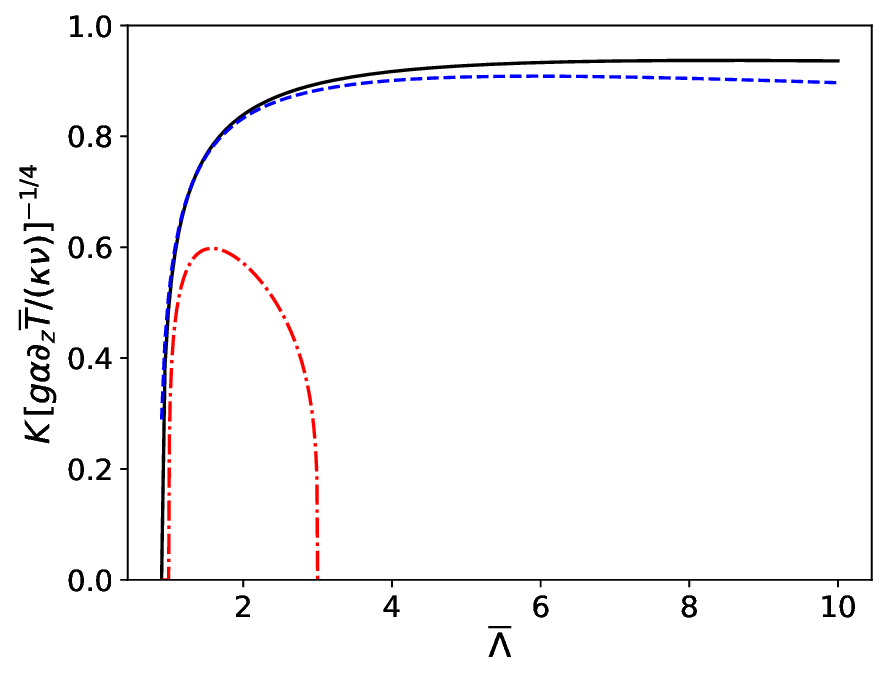}
\caption{$K [g \alpha \partial_z \overline{T} / (\kappa \nu)]^{-1/4}$ as a
function of $\overline{\Lambda}$ for the electrolyte investigated in this paper
($Pr=9$, $Sc=2000$, black continuous line), sea water ($Pr=7$, $Sc=700$, blue
dashed line) and salt and sugar in aqueous solution ($Pr=700$, $Sc=2100$, red dot
dashed line).}
\label{fig:prefactor_Lambda}
\end{figure}

The equations (\ref{eq:NSE}-\ref{eq:c}) linearized around $v_z=0$, $\theta=0$
and $s=0$ become
\begin{equation}
\partial_t v_z = - \frac{1}{\rho} \partial_z p + \nu \nabla^2 v_z + g \alpha
\theta - g \beta s
\end{equation}
\begin{equation}
\partial_t \theta + v_z \partial_z \overline{T} = \kappa \nabla^2 \theta
\end{equation}
\begin{equation}
\partial_t s + v_z \partial_z \overline{c} = D \nabla^2 s.
\end{equation}
These equations are  solved in an infinite domain with constant 
$\partial_z \overline{T}$ and $\partial_z \overline{c}$ by solutions of the
form
$(v_z,\theta,s) = (v_0,\theta_0,s_0) e^{\sigma t} e^{i(k_x x + k_y y)}$
to yield a growth rate $\sigma$ as a function of wavenumber
$k=\sqrt{k_x^2+k_y^2}$. One can then determine the wavenumber $K$ at which
$\sigma$ is maximum. The result of this computation is well known
\cite{Schmitt78,Schmit11} and is sometimes summarized as 
$K [g \alpha \partial_z \overline{T} / (\kappa \nu)]^{-1/4} = \mathrm{const}$.
However, there is also a dependence on $Pr$ and $Sc$ and also the density ratio
which in the unbounded model is best defined as
\begin{equation}
\overline{\Lambda} = \frac{\alpha \partial_z \overline{T}}{\beta \partial_z
\overline{c}}.
\end{equation}
There are no growing modes for $\overline{\Lambda} > Sc / Pr$. One naively
expects that narrow finger modes are replaced by overturning motion of
arbitrarily long wavelength for $\overline{\Lambda} < 1$, but this is
not strictly true as finger modes also exist for $\overline{\Lambda}$ slightly
below 1 for the $Sc$ and $Pr$ of interest to laboratory experiments
\cite{Schmit11}. Fig. \ref{fig:prefactor_Lambda} shows 
$K [g \alpha \partial_z \overline{T} / (\kappa \nu)]^{-1/4}$ as a function of
$\overline{\Lambda}$ for the most common laboratory fluids. One recognizes that
for cupric sulfate in sulfuric acid and for sea water, the product
$K [g \alpha \partial_z \overline{T} / (\kappa \nu)]^{-1/4}$ is close to $0.9$
for a wide range of $\overline{\Lambda}$.

\begin{figure}
\includegraphics[width=10cm]{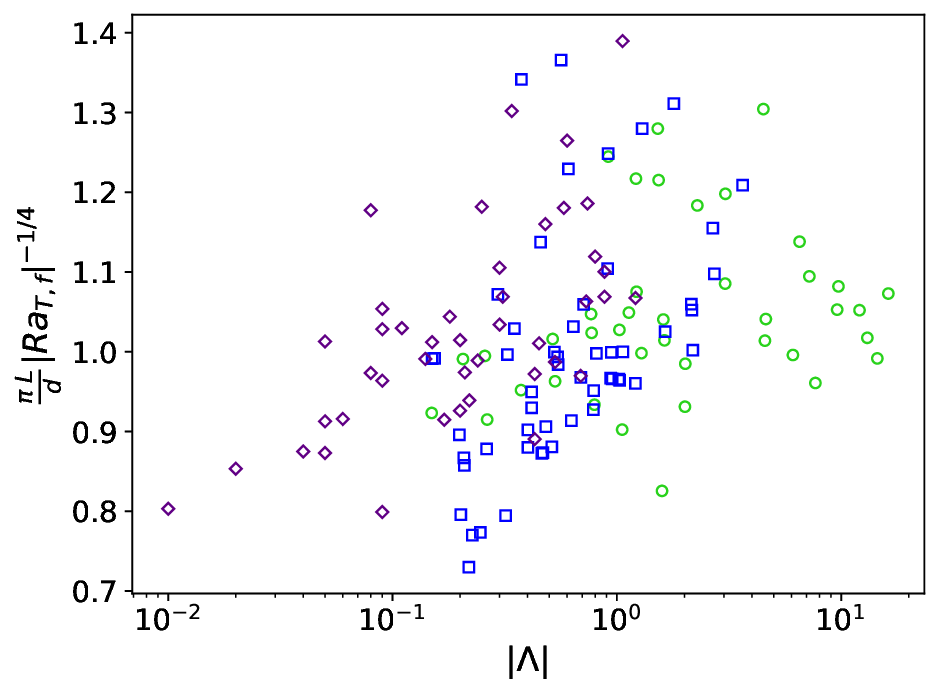}
\caption{$\frac{\pi L}{d} |Ra_{T,f}|^{-1/4}$ as a function of $|\Lambda|$. The symbols are the same as in fig. 
\ref{fig:param_space}.}
\label{fig:fastest_mode}
\end{figure}

We do not know from measurements the correct value of $\partial_z \overline{T}$
within the electrochemical cell. As a first guess, let us assume that 
$\partial_z \overline{T} = - \Delta T /L$. The wavenumber $K$ corresponds to
$\pi/d$, so that 
$K [g \alpha \partial_z \overline{T} / (\kappa \nu)]^{-1/4} =
\pi L / d |Ra_T|^{-1/4}$, which is shown in in Fig. \ref{fig:fastest_mode} as a
function of $|\Lambda|$. The points scatter around $1.1$ rather than $0.9$, but
apart from this 20\% discrepancy, the fastest growing mode assuming an interior temperature gradient of
$- \Delta T / L$ is a good prediction for the actual finger thickness.

We will now discuss by how much the true value of $\partial_z \overline{T}$ can
depart form $- \Delta T / L$. To this end, we first relate the fluctuations
$\theta$ to the mean $\overline{T}$ by
\begin{equation}
v_z \partial_z \overline{T} = \kappa (\partial^2_x + \partial _y^2) \theta.
\label{eq:elevator_T}
\end{equation}
This equation describes an elevator mode and follows from eq. (\ref{eq:T}) if
the flow is stationary, if $|v_x| \ll |v_z|$ (which is satisfied because of the
large anisotropy of velocity within the fingers \cite{Kellne14}), and if
$\theta$ does not depend on $z$. Let us check the self consistency of this last
assumption. Eq. (\ref{eq:elevator_T}) admits solutions of the form 
$v_z=v_0 \cos(\pi x/d)$ and 
$\theta = - \frac{v_0 d^2}{\pi^2 \kappa} \partial_z \overline{T} \cos(\pi x /
d)$ if $ \partial_z \overline{T}$ is constant, so that the total temperature
\begin{equation}
T = z \partial_z \overline{T} + \theta =
\partial_z \overline{T} \left[ z - \frac{v_0 d^2}{\pi^2 \kappa} \partial_z \overline{T} 
\cos(\frac{\pi x}{d}) \right]
\label{eq:T_solution}
\end{equation}
is constant on $z-\frac{v_0 d^2}{\pi^2 \kappa} \cos(\pi x / d)=\mathrm{const}$.
Isotherms are thus sinusoids with amplitudes $\Delta z$ given by
\begin{equation}
\frac{\Delta z}{L} = \frac{1}{\sqrt{2} \pi^2} \left( \frac{d}{L} \right)^2 Re Pr
\end{equation}
with 
$Re=\frac{L}{\nu} \left( \overline{v_0^2 \cos^2 (\pi x /d)} \right)^{1/2} = v_0 L /
(\sqrt{2} \nu)$.

\begin{figure}
\includegraphics[width=10cm]{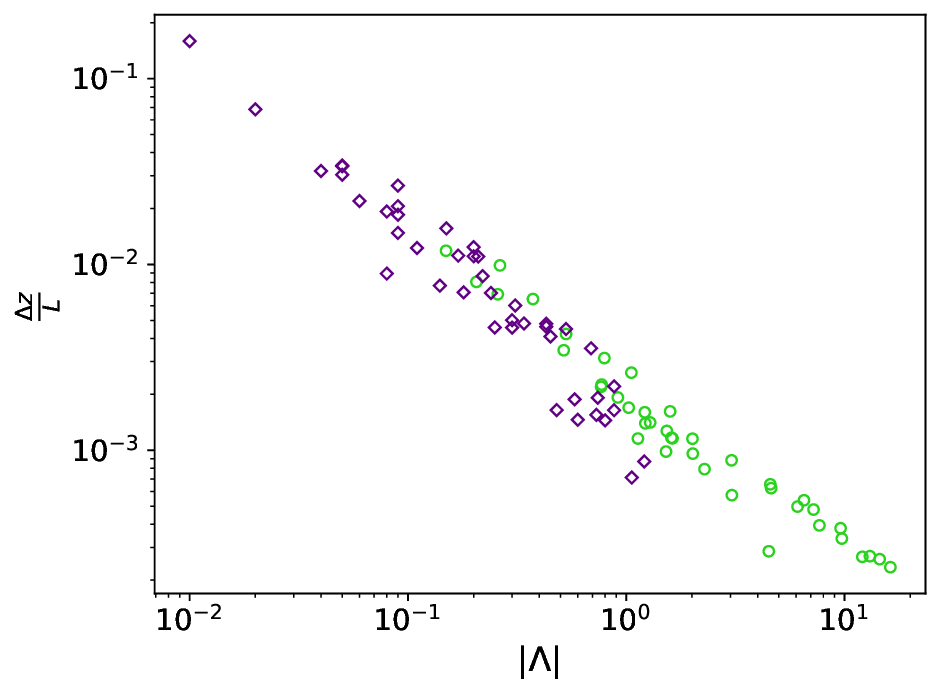}
\caption{$\frac{\Delta z}{L}$ as a function of $|\Lambda|$. The symbols are the same as in fig. 
\ref{fig:param_space}.}
\label{fig:Delta_z}
\end{figure}

The amplitude $\Delta z / L$ is shown in Fig. \ref{fig:Delta_z}. When $\Delta z
/ L$ is small, then the temperature field (\ref{eq:T_solution}) is compatible with the
boundary conditions in most of the volume. Deviations are expected only in
layers of size $\Delta z$ next to each boundary. Since $\Delta z / L < 0.1$ for
all points except one, we conclude that this solution provides us with a good
description of the temperature field in the cell. Note that the analogous
conclusion does not hold for the concentration field since the deformation
amplitude of the isopycnics is larger by the factor $Sc/Pr \approx 200$ and is
at least comparable with the cell height for $|\Lambda|<1$.

\begin{figure}
\includegraphics[width=10cm]{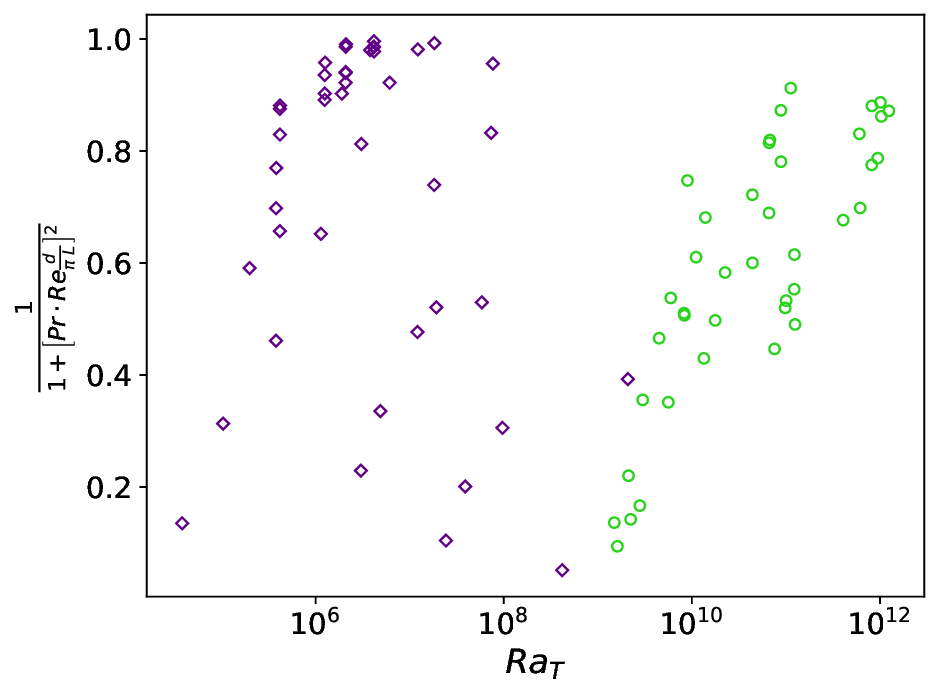}
\caption{The smallest possible value of $L T_z / |\Delta T|$ according to eq.
(\ref{eq:L_Tz_T}) is given by $1/[1+[Pr Re d / (\pi L)]^2]$ and is shown as a
function of $|Ra_T|$. The symbols are the same as in fig. \ref{fig:param_space}.}
\label{fig:L_dzT_DeltaT}
\end{figure}

We now proceed with the solution (\ref{eq:T_solution}) to compute the advective
heat transport as
\begin{equation}
\langle v_z \theta \rangle =
- \frac{v_0 d^2}{2 \pi^2 \kappa} \partial_z \overline{T}
\label{eq:vz_theta}
\end{equation}
For a piecewise linear profile of mean temperature with a vertical gradient
$T_z$ in the bulk and a gradient of $\Delta T_b / \lambda_T$ in thermal boundary
layers of thickness $\lambda_T$ and a temperature drop $\Delta T_b$ across them,
the gradients need to fulfill two conditions: The total temperature difference
across the cell needs to be $|\Delta T|$, 
\begin{equation}
2 \Delta T_b + T_z (L-2 \lambda_T) = |\Delta T|
\end{equation}
and the heat flux through the boundary layers must equal the flux through the
bulk, which leads with the help of eq. (\ref{eq:vz_theta}) to
\begin{equation}
-\kappa \frac{\Delta T_b}{\lambda_T} = - \kappa T_z - \frac{1}{2 \pi^2}
\frac{v_0^2 d^2}{\kappa} T_z.
\end{equation}
One can eliminate $\Delta T_b$ from these two equations to obtain
\begin{equation}
\frac{L T_z}{|\Delta T|} =
\frac{1}{1+ \frac{2}{\pi^2} \frac{\lambda_T}{L} \left( Pr Re \frac{d}{L}
\right)^2}.
\label{eq:L_Tz_T}
\end{equation}
If the actual finger thickness is equal to half the wavelength of the fastest growing
mode in an infinite domain with temperature stratification $T_z$, then
$\frac{\pi}{d} \left( \frac{g \alpha T_z}{\kappa \nu} \right)^{-1/4}
= \pi \frac{L}{d} |Ra_T|^{-1/4} \left( \frac{L T_z}{|\Delta T|} \right)^{-1/4}
=\mathrm{const}$.
This is different from the comparison in Fig. \ref{fig:fastest_mode} because of
the factor $(L T_z / |\Delta T|)^{-1/4}$. We need to know $\lambda_T$ to compute 
$L T_z / |\Delta T|$ from eq. (\ref{eq:L_Tz_T}), but we cannot measure
$\lambda_T$ directly. However, $2 \lambda_T / L \le 1$ by definition which puts a
lower bound on the possible values of $L T_z / |\Delta T|$. By evaluating eq.
(\ref{eq:L_Tz_T}) with $2 \lambda_T / L = 1$, one finds for most points
(see fig. \ref{fig:L_dzT_DeltaT}) that
$1 > L T_z / |\Delta T| > 0.2$
and hence
$1 > (L T_z / |\Delta T|)^{1/4} > 0.67$. In the worst case, the points near
$\pi \frac{L}{d} |Ra_T|^{-1/4}=1.4$ in Fig. \ref{fig:fastest_mode} have
$\frac{\pi}{d} \left( \frac{g \alpha T_z}{\kappa \nu} \right)^{-1/4}
=1.4/0.67=2.1$ which is a factor of $2.3$ above the value of $0.9$ expected for
most density ratios according to Fig. \ref{fig:prefactor_Lambda}. This means
that the observed finger thickness certainly agrees with the half wavelength of the fastest
growing mode to better than a factor $2.3$. If one accepts that the temperature
gradient is nearly unaffected by finger convection and equal to $-\Delta T/L$,
the agreement is to within a factor of $1.4/0.9 \approx 1.5$.
Fig. \ref{fig:fastest_mode} includes data for finger layers within staircases,
so that the same conclusion holds for this class of fingers and if one assumes a
thermal gradient of $-\Delta T_f/L_f$ within the fingers.
Considering the scatter in Fig. \ref{fig:L_dzT_DeltaT} it does not seem to be
worthwhile to improve these factors. At any rate, it is difficult to find a
better and trustworthy alternative to the gross overestimate of $\lambda_T$ in
Fig. \ref{fig:L_dzT_DeltaT}. For example, one may attempt to model the diffusion
in the temperature boundary layer, but the next section will point at
limitations of this approach.

\begin{figure}
\includegraphics[width=10cm]{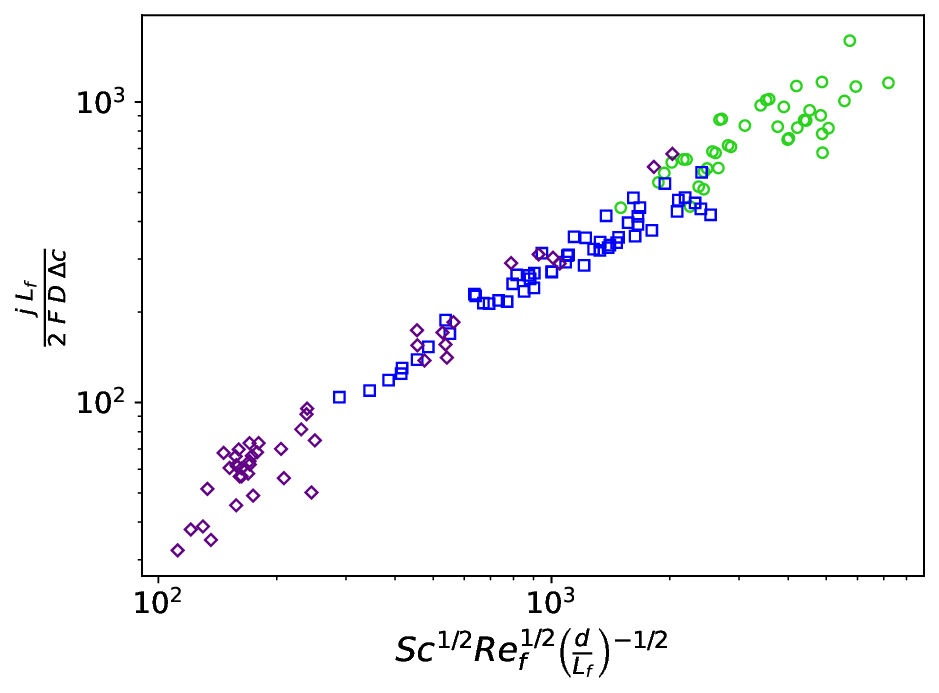}
\caption{$j L_f / (2F D \Delta c)$, which is equal to $Sh$ for fingers extending
through the whole cell, as a function of $(Sc Re_f \frac{L_f}{d} )^{1/2}$.
The symbols are the same as in fig. \ref{fig:param_space}.}
\label{fig:Sh_bl}
\end{figure}

\section{Salinity flux in fingers}
\label{section:flux_fingers}

The simplest model for the boundary layers of the concentration field near one
of the solid boundaries is that of ions diffusing with diffusivity $D$ into a
flow parallel to the boundary of velocity $v$. It takes the time $d/v$ for the
flow to travel horizontally across a finger so that the boundary layer thickness
$\lambda_c$ is proportional to $(D d/v)^{1/2}$. Let the concentration drop
across each boundary layer be $\Delta c_b$, so that the Sherwood number is given
by $Sh = L \Delta c_b / ( \lambda_c \Delta c)$. If there is no gradient left in
the interior of the flow, then $\Delta c_b = \Delta c / 2$ and one obtains the
classical formula $Sh = L / (2 \lambda_c)$. This leads with 
$\lambda_c \propto (D d/v)^{1/2}$ to
\begin{equation}
Sh \propto (Sc Re \frac{L}{d} )^{1/2}.
\label{eq:Sh_bl}
\end{equation}
This relation is verified in Fig. \ref{fig:Sh_bl}. The good agreement found in
this figure does not prove that $\Delta c / \Delta c_b = 2$, but it shows that
$\Delta c / \Delta c_b$ varies little in dependence of the Rayleigh numbers.
Salinity cannot diffuse from one finger to the neighboring finger during the
time it takes to travel vertically across the cell at finger velocity
\cite{Hage10}. Fingers therefore maintain the ion concentration they had when
they started from one boundary until they hit the opposite boundary. Because of
the top bottom symmetry of the system, the horizontally averaged concentration
must therefore be $c_0$ everywhere and the gradient $\partial_z \overline{c}$ is
zero in the bulk. Even though we cannot measure $\Delta c_b$ directly, it is
very plausible that $\Delta c / \Delta c_b = 2$.

The data points at low $Ra_c$ in Fig. \ref{fig:param_space} obey simple scaling
laws \cite{Hage10} for $Re$, $Sh$ and $d/L$. It is remarkable that these scaling
laws exactly verify $Sh \propto (Re \frac{L}{d} )^{1/2}$ even though the
exponents in these scalings were obtained from three independent fits to
experimental data.

\begin{figure}
\includegraphics[width=10cm]{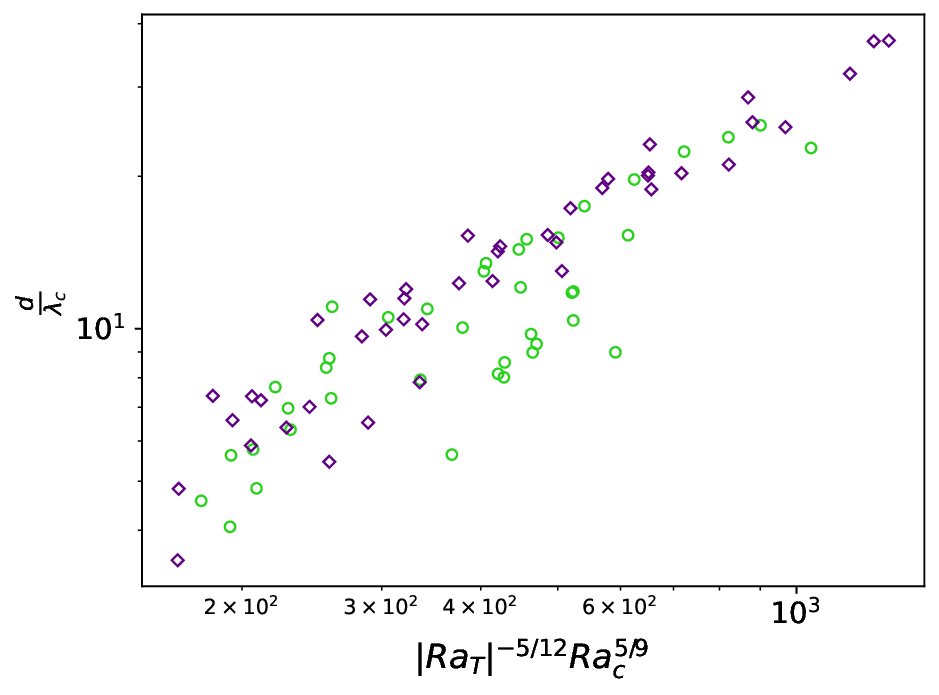}
\caption{$d/\lambda_c$ as a function of $|Ra_T|^{-5/12} Ra_c^{5/9}$ which is the
scaling for $d/\lambda_c$ according to Ref. \onlinecite{Hage10}. The symbols are the same as in fig. 
\ref{fig:param_space}.}
\label{fig:d_lambda_c}
\end{figure}

The picture of ions diffusing into a parallel flow to form a boundary layer is
only useful if the boundary layer thickness $\lambda_c$ is small compared with
the horizontal distance over which the parallel flow exists, which is the finger
width $d$ in our application. If on the contrary $d/\lambda_c \gg 1$ is not satisfied, one must
expect significant ion diffusion into the stagnation points that exist on the
boundaries at the edges of fingers and near which the flow is not parallel to the
boundary. Fig. \ref{fig:d_lambda_c} shows that for $\lambda_c$ computed from 
$\lambda_c = L / (2 Sh)$, one indeed finds $d/\lambda_c > 3.5$ for all data
points and $d/\lambda_c > 10$ for the majority of them.

The temperature boundary layer thickness $\lambda_T$ is according to the same
reasoning given by $\lambda_T \propto (\kappa d/v)^{1/2}$ and
$\lambda_T = (Sc/Pr)^{1/2} \lambda_c \approx 14 \lambda_c$ so that 
$d/\lambda_T < 1$ for many of the data points in our collection and 
$\lambda_T \propto (\kappa d/v)^{1/2}$ must be a poor approximation of the true
temperature boundary layer thickness. The model of a boundary layer created by
diffusion into parallel shear flow is applicable to the salinity boundary layer and
was already successfully used in some treatments of double diffusive convection
\cite{Yang15, Tassin24} but it cannot be used throughout the full data set
considered here for the temperature boundary layer.

Most previous experiments have confirmed the so called 4/3-law which states that
$j \propto \Delta c^{4/3}$, or in non dimensional form including all control
parameters, $Sh \propto f(Pr,Sc,\Lambda) Ra_c^{1/3}$ with an undetermined
function $f$. The exponent $1/3$ is not the preferred value in the fits to the
data obtained from the electrochemical cell, but it was already noted in Ref.
\onlinecite{Hage10} that the data are nearly indistinguishable from and hence compatible with
the 4/3-law. The point that deserves attention here is not so much the exponent
as the prefactor.

\begin{figure}
\includegraphics[width=10cm]{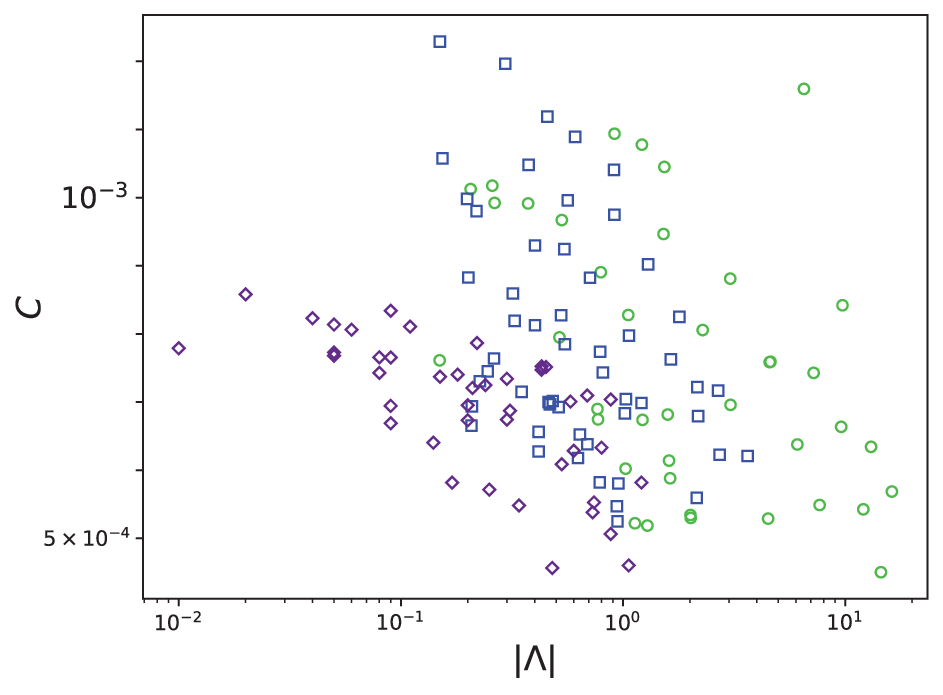}
\caption{$C = \beta j / [ 2 F (g \kappa)^{1/3} (\beta \Delta c)^{4/3}]$ as a
function of $|\Lambda|$. The symbols are the same as in fig. \ref{fig:param_space}.}
\label{fig:Sh_4_3}
\end{figure}

For better comparison with other experiments, let us consider the ion or
salinity flux $F_s$ in $mol/ (m^2 s)$ which in the electrochemical cell can be
determined from eq. (\ref{eq:def_Sh}) as
$F_s = j / (2 F) = Sh D \Delta c / L$
and seek the proportionality constant $C$ in
\begin{equation}
\beta F_s = C (g \kappa)^{1/3} (\beta \Delta c)^{4/3}.
\label{eq:C}
\end{equation}
Values of $C$ for finger convection in sea water can be found in Fig. 7 of
\citet{Schmitt79} who found that $C$ decreases with $|\Lambda|$ at small
$|\Lambda|$ and is approximately constant beyond $|\Lambda|=3$ with
$C \approx 0.05$. A compilation of data appears in Fig. 5 of \citet{Taylor89}
and shows compatible numbers. However, the values of $C$ for the electrochemical
cell shown in Fig. \ref{fig:Sh_4_3} lie in the interval
$5 \times 10^{-4} \lesssim C \lesssim 1.2 \times 10^{-3}$ and are therefore smaller by
approximately a factor 100.

It is unlikely that such a large factor is due to minor changes in $Pr$ and $Sc$
with $Pr=7$ and $Sc=700$ for sea water and $Pr=9$ and $Sc \approx 2000$ in 
our electrolyte. The ranges of Rayleigh numbers are also broadly similar in
Fig. 7 of \citet{Schmitt79} and Fig. \ref{fig:param_space}. The potentially relevant
difference that comes to mind is the nature of the finger tips. In the
electrochemical cell, the fingers have one tip ending on a solid wall if they
are part of a staircase, and both tips if they cross the entire cell. Older
experiments started from two layers on top of each other, separated either by a
steep step or a smooth profile of salinity and temperature, but in all cases,
fingers were not in contact with a solid wall. It is plausible that this gives
more freedom to the transport into and out of the fingers at their ends, even
though the amplitude of the effect remains surprising.

An early experiment on double diffusive finger convection in salty water
\cite{Turner67} measured the salinity flux between two layers, a warm fresh
layer superimposed on cold salty water, with density ratios between 2 and 10. By
lack of any other data, the author compared this salinity flux with the well
investigated heat flux in thermal Rayleigh-B\'enard convection and found a
discrepancy by a factor of 10 to 40 between the two fluxes. Even though this
study had to compare a salinity flux with a heat flux, it already indicated a
large difference of fluxes depending on whether a solid wall is present or not.

Values of $C$ can also be obtained from the 2D computations for $Pr=7$ and
$Sc=700$ by \citet{Li23} after rewriting eq. (\ref{eq:C}) as
$C = Ra_c^{-1/3} Sc^{-2/3} Pr^{1/3} Sh$. 
These computations simulate fluid in a plane layer,
but visualizations indicate that fingers are not necessarily in contact with one
of the boundaries. One can compute from the tabulated results in the range 
$10^9 \le Ra_c \le 10^{12}$ that $0.0012 \le C \le 0.0015$ at $|\Lambda|=1$
and $0.04 \le C \le 0.06$ at $|\Lambda|=3$. These results are unusual in that $C$
increases with $|\Lambda|$. The numerically obtained $C$ lie in between the
results in \citet{Schmitt79} and \citet{Taylor89} and ours and do not provide us
with a definitive answer on why the $C$ in the electrochemical cell is so small. 

A related set of results was obtained for the sugar/salt system. In these
experiments in aqueous solution, sugar plays the role of the slowly diffusing
and salt that of
the rapidly diffusing component. The ratio of diffusivities is only 3 for this
combination. Comparing again the flux law for the flux between two layers
measured experimentally \cite{Taylor96} with the flux computed numerically in a
plane layer crossed by fingers, \citet{Radko00} found good agreement indicating
no difference between the dynamics of a bounded finger layer and fingers in a
two layer system. The flux reducing effect of the solid wall seems to depend on
parameters that remain to be identified.

\section{Salinity flux in staircases}
\label{section:flux_staircases}

The staircases were mostly omitted from the discussion so far because we do not
have access to all their control parameters or their Sherwood number. We do not
know the amplitude of the concentration
and temperature differences that drive the convection in the fingers
that are part of staircases. $\Delta T_f$ enters Fig.
\ref{fig:fastest_mode} only to the power $1/4$ so that the figure is rather
insensitive to the correct value of $\Delta T_f$ and it is innocuous to compare all
types of fingers in this figure.

There are two other relations which actually inform us about the correct
concentration difference for fingers in staircases. The first observation 
concerns eq. (\ref{eq:Sh_bl}).
Fig. \ref{fig:Sh_bl} was drawn with a $\Delta c$ in the definition of $Sh$ 
which is by definition
the correct choice for fingers crossing the whole cell, but as can be seen in
the figure, it is also the correct choice for fingers within staircases! In the
derivation of eq. (\ref{eq:Sh_bl}), $\Delta c$ stands for twice
the concentration drop across the boundary layer. According to Fig.
\ref{fig:Sh_bl}, this quantity is 
identical to $\Delta c$, the concentration difference between the two
electrodes, for both fingers in staircases and fingers spanning the whole cell.
This then implies that the horizontally averaged concentration of ions outside the boundary layers is
the same in both types of fingers. This means that the concentration contrast
between rising and sinking fingers must always be $\Delta c$ to obtain the
correct horizontal average. A finger falling down from the top plate must meet
with fingers carrying the ion concentration of the bottom plate to yield $c_0$
in the horizontal average. This implies
that a blob of fluid leaving for example the bottom plate with a low ion
concentration rises through both the lower finger layer and the central
convecting layer without gaining ions so that it arrives at the bottom edge of
the top layer with still the same ion concentration. This applies a concentration
difference of a full $\Delta c$ to the top layer. The symmetric reasoning holds
for a blob leaving the top plate passing unscathed both the top layer and the
central layer so that again there is a concentration difference of $\Delta c$
applied to the bottom layer, too. It was expected that fluid travels through the
finger layers without a change in ion concentration \cite{Hage10}, but it is more surprising to
see that the central layer does not mix the fluid and does not equalize the
concentration in its interior. For this to be possible, the blobs which have
traversed the central layer and impinge on the edge of a finger layer must be quickly
absorbed by that layer before they get entrained in the circulation of the
central convection roll which would inevitably lead to mixing. The interface
between finger layers and convection rolls must be very permeable.

This picture is supported by the behaviour of the proportionality constant $C$
in the 4/3-law. The expression for $C$ in eq. (\ref{eq:C}) also contains a
$\Delta c$, which is identical for all classes of fingers in Fig.
\ref{fig:Sh_4_3}. This figure shows that both fingers extending through the
whole cell and fingers in staircases obey the 4/3-law with the same
proportionality constant if and only if it is again assumed that the
concentration difference across every finger layer is $\Delta c$.

These observations motivate the definition of $Ra_{c,f}$ in eq. (\ref{eq:Ra_f})
with $\Delta c_f = \Delta c$.
The relations represented in Figs. \ref{fig:Sh_bl} and \ref{fig:Sh_4_3} are
independent of the temperature field so that they probe the correct $\Delta c_f$
independently of the correct choice for $\Delta T_f$.
The temperature difference $\Delta T_f$ in the
definition of $Ra_{T,f}$ is chosen to be $\Delta T_f=\Delta T/2$
because this leads to the best fit in Fig. \ref{fig:d_L}. It is possible that
because of the more rapid diffusion of heat, temperature differences in the
central convecting layer are erased while concentration differences remain, so
that half of the total temperature drop occurs over each finger layer. However,
$Ra_{T,f}$ only occurs in Figs. \ref{fig:d_L} and \ref{fig:fastest_mode}, raised
to the power 1/3 or 1/4, so that the data do not discriminate well between
different choices of $\Delta T_f$.

\section{Conclusion}

Measurements on double diffusive finger convection in an electrochemical cell
were analyzed in this paper and compared with results obtained from other double
diffusive systems. A major motivation for this comparison is that fingers in the
electrochemical cell at the control parameters investigated so far are always
limited in their extent by a solid boundary. This is different from other
experimental systems in which fingers are detached from solid boundaries.

A generalization of the scaling law \cite{Hage10} found for finger thickness at
$Ra_c < 10^{12}$ to all fingers on which measurements are available in the
electrochemical cell leads to a satisfactory fit to all data, including to those
for fingers within staircases. However, there is no ab initio theory to derive
this scaling law. Past experiments and simulations indicate that finger widths
agree well with half the wavelength of the fastest growing mode found in linear
stability analysis. There is no theory either for why the flow in the saturated
state should have the same finger size as during a growth phase, but it is
intuitive to assume so. For the double diffusive convection in the
electrochemical cell, the observed finger width and the width of the fastest
growing mode agree to within a factor $1.5$ if one assumes that the thermal
gradient along the fingers is the average thermal gradient imposed by boundary
conditions. The actual gradient within the fingers might be less, but the
agreement must still be to within a factor $2.3$ or better.

The salinity transport conforms with the 4/3-law, but the proportionality
factor appearing in this law is much smaller in the experiments analyzed here
than in previous experiments. A plausible explanation for this discrepancy is
precisely that fingers were surrounded by fluid from all sides in the older
experiments, whereas at least
one tip of every finger borders a solid wall in the experiments in the
electrochemical cell. The presence of a wall certainly restricts the dynamics
and reduces the transport through the finger tips, but it is noteworthy that
this reduction is by two orders of magnitude.

The transport through finger tips in the absence of a solid boundary is so
efficient that it prevents overturning layers sandwiched between two finger
layers in staircases to mix the fluid in their interior well enough to
annihilate concentration gradients. Plumes colliding with finger tips seem to be
siphoned into the finger layer before they loose much concentration contrast
with their environment. Some mixing in the overturning layer must of course
occur, but it is too small to be detectable in staircases that contain just one
overturning layer. The implications for staircases with many more layers remain
to be explored.

Data sharing is not applicable to this article as no new data were created or analyzed in this study.

\begin{acknowledgments}
This work was supported by the DFG under the grant Ti243/15 as part of the
priority programme ''Reconstructing the Deep Dynamics of Planet Earth over
Geologic Time''.
\end{acknowledgments}


%

\end{document}